\documentclass[aps,prb,twocolumn,showpacs,groupedaddress]{revtex4}
\usepackage{graphicx}

\begin{document}

\title{Effect of pressure on the superconducting and spin-density-wave states of Sm(O$_{1-x}$F$_x$)FeAs}
\author{B. Lorenz$^1$, K. Sasmal$^1$, R. P. Chaudhury$^{1}$, X. H. Chen$^2$, R. H. Liu$^2$, T. Wu$^2$, and C. W. Chu$^{1,3,4}$}
\affiliation{$^{1}$Department of Physics and TCSUH, University of Houston, Houston, TX 77204, USA} \affiliation{$^{2}$Hefei National Laboratory
for Physical Science at Microscale and Department of Physics, University of Science and Technology of China, Hefei, Anhui 230026, People's
Republic of China} \affiliation{$^{3}$Lawrence Berkeley National Laboratory, 1 Cyclotron Road, Berkeley, CA 94720, USA} \affiliation{$^{4}$Hong
Kong University of Science and Technology, Hong Kong, China}
\date{\today }

\begin{abstract}
High temperature superconductors with a T$_c$ above 40 K have been found to be strongly correlated electron systems and to have a layered
structure. Guided by these rules, Kamihara et al. discovered a T$_c$ up to 26 K in the layered La(O$_{1-x}$F$_x$)FeAs. By replacing La with
tri-valence rare-earth elements RE of smaller ionic radii, T$_c$ has subsequently been raised to 41 - 52 K. Many theoretical models have been
proposed emphasizing the important magnetic origin of superconductivity in this compound system and a possible further T$_c$-enhancement in
RE(O$_{1-x}$F$_x$)FeAs by compression. This later prediction appears to be supported by the pressure-induced T$_c$-increase in
La(O$_{0.89}$F$_{0.11}$)FeAs observed. Here we show that, in contrast to previous expectations, pressure can either suppress or enhance T$_c$,
depending on the doping level, suggesting that a T$_c$ exceeding 50's K may be found only in the yet-to-be discovered compound systems related
to but different from RE(O$_{1-x}$F$_x$)FeAs and that the T$_c$ of La(O$_{1-x}$F$_x$)FeAs and Sm(O$_{1-x}$F$_x$)FeAs may be further raised to
50's K.
\end{abstract}

 \maketitle











There exists in nature a large class of equiatomic quaternary layered compounds REOTPn (RE = La, Nd, Sm, Gd; T = Mn, Fe, Co, Ni, Cu; Pn = P, As,
Sb) with a tetragonal structure of the ZrCuSiAs type.\cite{15} The rare-earth transition oxypnictides REOTPn consist of transition-metal
pnictide (TPn)-layers sandwiched by rare-earth oxide (REO)-layers. Similar to the cuprate high temperature superconductors, the charge carriers
are supposed to flow in the (TPn) layers while the (REO)-layers inject charge carriers to the former via the so-called "modulation doping" while
retaining the layer integrity of the (TPn) layers. However, the coordination structure of REOTPn is different from that of the high Tc cuprates:
the divalent T is tetrahedrally coordinated with four Pn ions, whereas the divalent Cu forms the four-fold square plane. Based on the above
analysis, superconductivity was first discovered in LaOFeP with a T$_c\sim$ 4 K.\cite{16} By increasing the carrier concentration through the
partial replacement of O by F, the T$_c$ was raised to $\sim$ 9 K. Shortly afterward, LaONiP was found to exhibit a T$_c\sim$ 3 K.\cite{17}
However, immense excitement did not arise until the very recent discovery of the 26 K superconductivity in the F-doped LaOFeAs.\cite{2} The
refined X-ray diffraction data show that the oxidation numbers of REOFeAs are RE$^{3+}$O$^{2-}$Fe$^{2+}$As$^{3-}$, where the conducting iron
arsenide (FeAs)$^{1-}$ layers are stacked alternately with the less conducting rare-earth (REO)$^{1+}$ layers.\cite{15}

Immediately after the discovery of a T$_c$ of 26 K in La(O$_{1-x}$F$_x$)FeAs,\cite{2} T$_c$ was drastically raised to 43 K in
Sm(O$_{1-x}$F$_x$)FeAs,\cite{3} followed by reports of a T$_c$ up to 41 K in Ce(O$_{1-x}$F$_x$)FeAs,\cite{4} 52 K in
Pr(O$_{1-x}$F$_x$)FeAs,\cite{5} and 50 K in Nd(O$_{1-x}$F$_x$)FeAs.\cite{6} These are the first instances that T$_c$'s above 40 K, the
theoretical T$_c$-limit prior to the discovery of the 93 K YBa$_2$Cu$_3$O$_7$ superconducting cuprate,\cite{18} have been found outside the
layered cuprate compound system. The recent discoveries have generated great enthusiasm about the future of high temperature superconductivity.
These Fe-based rare-earth oxyarsenides RE(O$_{1-x}$F$_x$)FeAs are expected to provide a new material base for studying the origin of high
temperature superconductivity and to offer a novel avenue to achieving superconductivity at a temperature surpassing the record T$_c$ of the
cuprates. Indeed, the crucial role of the magnetic Fe-element in the occurrence of the relatively high T$_c$ in these compounds is unexpected,
since the presence of magnetic ions tends to be antagonistic to the conventional s-wave superconductivity. Unconventional superconductivity has
been proposed by many.\cite{7,8,9,10,11,12,13,19,20} Much higher T$_c$ has also been suggested in this class of compounds by fine-tuning through
doping and/or applying pressure.\cite{11,12,14} The suggestion appears to be consistent with the initial observation of the T$_c$-increase of
La(O$_{1-x}$F$_x$)FeAs due to the possible internal pressure induced by the replacement of La by the smaller rare-earth elements.\cite{3,4,5,6}
It seems also to be corroborated by the T$_c$-enhancement of La(O$_{1-x}$F$_x$)FeAs by external pressures at a rate of dT$_c$/dP $\sim$ 1.2
K/GPa.\cite{14} This new compound system is expected to have a softer characteristic\cite{12} and, when an enhanced T$_c$ is achieved, may thus
alleviate some of the burdens in high temperature superconducting wire-fabrication encountered for the second-generation superconducting wires
that use YBa$_2$Cu$_3$O$_7$. However, it is not clear whether such a positive pressure effect on T$_c$ is true for all other superconducting
rare-earth Fe oxyarsenides, whether the rapid saturation of T$_c$ at $\sim$ 50's K with x reported is intrinsic, and whether pressure can
further raise the T$_c$ of those compounds with their T$_c$ already over 50's K.

Band calculations show that the electronic structure of REOFeAs is quasi-two-dimensional and semi-metal-like at the verge of instabilities,
suggesting the possible existence of different competing states against the superconducting state, such as spin-density-wave (SDW),
antiferromagnetism, and ferromagnetism.\cite{7,8,9,10} Indeed, magnetic, resistive, and optical measurements of REOFeAs display anomalies at
$\sim$ 150 K,\cite{2,9} indicative of the opening of a SDW gap on cooling and partial reduction of the Fermi surface due to Fermi surface
nesting between the electrons and holes.\cite{9} Doping through the partial replacement of O by F or application of external pressure is thus
suggested to narrow and eventually eliminate the SDW gap, leading to the appearance of superconductivity. This appears to be supported by the
experimental observations that superconductivity takes place as soon as the 150 K resistive anomaly is suppressed by F-doping. However, the
rapid rise of T$_c$ of RE(O$_{1-x}$F$_x$)FeAs to its x-insensitive maximum plateau is different from the cuprates and not yet understood. The
complexity in sample preparation of RE(O$_{1-x}$F$_x$)FeAs may be able to account, at least partially, for the observation. The exact effect of
SDW on superconductivity in these compounds remains unclear. Examining the pressure influence on SDW may provide insight into the relationship
between the two phenomena without chemical complications in doping. To address some of the questions raised above, we have chosen to investigate
the pressure effect on the Sm(O$_{1-x}$F$_x$)FeAs samples with nominal x = 0, 0.05, 0.13, and 0.3, covering the nonsuperconducting and the
superconducting regions.

All samples were prepared by solid state reaction with the precautions described previously.\cite{2,3} X-ray spectra show the typical
diffraction profile for Sm(O/F)FeAs (Fig. 1). Various impurity phases are detected in particular for x=0.3 as indicated in Fig. 1. The dc
magnetization for the superconducting samples was measured in a superconducting quantum interference device (SQUID) at ambient pressure. Samples
used for resistivity measurements had a typical size of 3 mm length and an area of 1 mm$^2$. The resistance was measured using the standard
four-lead technique. The electrical contacts were made by attaching platinum wires using silver paint. The contact resistance was of the order
of a few $\Omega$. The low-frequency (19 Hz) resistance bridge (LR700) was employed for resistivity measurements. High pressure measurements
were carried out employing the Be-Cu clamp method with the pressure determined by the Pb-pressure gauge.\cite{21} A mixture of Fluorinert 70 and
77 liquid was used as the pressure transmitting medium.

The magnetic susceptibility of the superconducting samples was
measured at ambient pressure in 5 Oe and the diamagnetic shielding
signal corresponds to $>$ 30\% of superconducting volume fraction
for both samples. This value is reasonable in view of the
nonsuperconducting impurity phases present in the samples. The
resistance (R) variations with temperature in our samples are in
agreement with previous reports. A R-maximum at T$_{SDW}$ indicative
of the onset of magnetic order (spin density wave, SDW) is evident
in the x = 0 and 0.05 nonsuperconducting samples with T$_{SDW}$
decreasing from 150 K to 125 K at ambient pressure, but not in the x
= 0.13 and 0.3 superconducting samples with an onset T$_c$
increasing from $\sim$30 K to $\sim$48 K (Fig. 2), consistent with
the F-doping effects expected from theoretical predictions.

Under pressure, the room temperature R of the x = 0.3 sample
decreases rapidly with initial pressure-increase and continues to
decrease, but only slowly, at higher pressures during the
pressure-increase cycle, perhaps due to an initial pressure-induced
compaction of the sample. However, pressure reduction results in an
irreversible R-increase, attributed to the pressure-induced
degradation of the sample. To define the superconducting T$_c$
consistently, we have taken the inflection point temperature of the
R-T curve or the peak-temperature of the dR/dT vs. T plot as our
T$_c$. A T$_c$ so-defined is expected to be lower than the
previously reported T$_c$'s, most of which referred to the onset
temperatures. The T$_c$ of the x = 0.3 sample is suppressed by
pressure at a rate of dT$_c$/dP $\sim$ - 2.3 K/GPa (Fig. 3), in
contrast to previous suggestions. It should be noted that the T$_c$
of 42.5 K upon the complete release of pressure is slightly lower
than the starting value, probably due to the combination effect of
sample degradation and the residual pressure locked in the high
pressure cell. On the other hand, for the x = 0.13 sample, T$_c$
increases with pressure from 24.7 K to 25.55 K at 0.94 GPa at a rate
of $\sim$ 0.9 K/GPa, and then stays at 25.55 K at higher pressures
(Fig. 4), while the room temperature R varies with pressure as in
the x = 0.3 sample. The T$_c$ after the complete release of pressure
became higher, attributable to the sample change and the residual
stress. Therefore, pressure effects on the T$_c$ of the two samples
examined are reversible except when the pressure is finally removed.
The latter can be due to the polycrystallinity and low density of
the samples which raises the possibility that the release of
pressure may weaken the interactions between grains. The unusual
pressure dependence of T$_c$ of this sample observed may be a
reflection of the fine electronic structure near the Fermi surface.
To study the behavior of the SDW state under pressure, we measured
only the x = 0.05 nonsuperconducting sample. It is evident that the
R-peak is suppressed by pressure (Fig. 5). For better definition, we
take the peak temperature of dR/dT as the T$_{SDW}$, and the
suppression of the SDW state by pressure is shown in Fig. 5.

In contrast to theoretical predictions that F-doping and pressure would have the same effect in suppressing the SDW state and enhancing the
superconducting state of Sm(O$_{1-x}$F$_x$)FeAs,\cite{9,12,14} we found that pressure can either promote or suppress the superconducting state,
depending on the doping level of x, whereas pressure always suppresses the SDW state. It has been shown that the T$_c$ of the cuprate high
temperature superconductors varies with carrier concentration (n) following a universal parabolic rule with T$_c$ peaks at a carrier
concentration n$_0$.\cite{22} T$_c$ increases with n in the so-called underdoped region where n$<$n$_0$, but deceases with n when the compound
is in the so-called overdoped region where n$>$n$_0$. It has also been demonstrated\cite{23} that dT$_c$/dP is negative when n$\geq$n$_0$ and
positive when n$<$n$_0$. The Sm(O/F)FeAs system seems to be similar to the high-T$_c$ cuprates although the current doping levels achieved do
not reveal the expected decrease of T$_c$ yet.\cite{27} Further increase of the electron number in the FeAs layers by improved methods of doping
should therefore result in a decrease of T$_c$. With this conjecture in mind, one can conclude that the x = 0.30 sample with a negative
dT$_c$/dP must lie close to the overdoped region and the x = 0.13 sample with a positive dT$_c$/dP in the underdoped region. This suggests that
the T$_c$ of Sm(O$_{1-x}$F$_x$)FeAs should peak between nominal x = 0.13 and 0.3 and within a T$_c$ range of 50 - 60 K. Indeed, a systematic
control of x in Sm(O$_{1-x}$F$_x$)FeAs has just led to an enhanced T$_c$ of 53 K. Furthermore, in the cuprate high temperature superconductors
REBa$_2$Cu$_3$O$_7$ (REBCO), where RE = Y and rare earth elements, RE controls the stability of the crystal structure but is electronically
isolated from the superconductivity of the compound, because the density of states of RE lies deep below the Fermi level.\cite{24} Therefore,
the T$_c$ of REBCO has been observed to vary with n universally and the maximum T$_c$ falls into the narrow range of 90's K independent of RE.
Similarities between the layered structures of RE(O$_{1-x}$F$_x$)FeAs and REBCO led us to conjecture that, like REBCO, RE(O$_{1-x}$F$_x$)FeAs
will have a similar T$_c$-variation with x and a narrow maximum T$_c$ range in $\sim$ 50's K for all RE, provided that the ZrCuSiAs layered
structure can be stabilized, in spite of the great variation from 26 to 52 K of the maximum T$_c$ of RE(O$_{1-x}$F$_x$)FeAs reported for
different RE's.\cite{2,3,4,5,6} A systematic study on the doping effect is warranted and should yield a non-monotonic T$_c$-x relation instead
of those previously reported. To increase the maximum T$_c$ of La(O$_{1-x}$F$_x$)FeAs from 26 K to 50's K is a strong possibility. On the other
hand, several cuprate high T$_c$ compound systems similar to REBCO exist with T$_c$ up to 134 K at ambient\cite{25} and 164 K at 30
GPa.\cite{26} It is not unlikely that T$_c$ above 50's K will be found in a yet-to-be-discovered compound system similar to but different from
REOFeAs with proper doping.

The suppression of the SDW state is clearly evidenced by the shifting of the resistance peak at T$_{SDW}$ to a lower temperature by F-doping and
also by the diminishing of the resistance peak near T$_{SDW}$ upon the application of pressure, in general agreement with the band calculations
(Fig. 5). Questions of whether the appearance of superconductivity requires the complete disruption of the SDW gap, as many of the published
results suggest, and if a direct interaction exists between the superconducting and the SDW states remain unanswered. A systematic high pressure
study on compounds very close to the border between the superconducting and SDW states should help address these questions.

\begin{acknowledgments}
Discussions with Y. Y. Xue are greatly appreciated. The work in Houston is supported in part by the T. L. L. Temple Foundation, the John J. and
Rebecca Moores Endowment, the United States Air Force Office of Scientific Research, and the State of Texas through the TCSUH. The work in China
is supported in part by the Nature Science Foundation of China and by the Ministry of Science and Technology of China.
\end{acknowledgments}

\bibliographystyle{phpf}

\begin{figure}
\caption{(Color online) X-ray spectra of SmO$_{1-x}$F$_x$FeAs for
x=0, 0.05, 0.13, 0.3 (bottom to top). Several impurity phases are
indicated by different symbols.} \caption{(Color online) R vs. T of
Sm(O$_{1-x}$F$_x$)FeAs at various doping levels. The doping levels
are noted next to the curves.} \caption{(Color online) T$_c$ vs. P
of Sm(O$_{0.7}$F$_{0.3}$)FeAs. The numbers denote the sequential
order of the experimental runs. Lower left inset: R vs. T at P=0 GPa
(1), 0.94 (2), 1.52 (3), 0.66 (4), $\sim$ 0+ (5). Upper right inset:
R' = dR/dT vs. T.} \caption{(Color online) T$_c$ vs. P of
Sm(O$_{0.87}$F$_{0.13}$)FeAs. The numbers denote the sequential
order of the experimental runs. Lower left inset: R vs. T at P=0 GPa
(1), 0.94 (2), 1.52 (3), 0.66 (4), $\sim$ 0+ (5). Upper right inset:
R' = dR/dT vs. T.} \caption{(Color online) R vs. T for
Sm(O$_{1-x}$F$_x$)FeAs at various doping levels and pressures. x=0
at P=0 GPa (dashed line); x=0.05 at P=0 GPa (open circles), 0.94
(solid circles), 1.51 (open triangles). Inset: R'=dR/dt vs. T for
the data shown in the main panel.}
\end{figure}

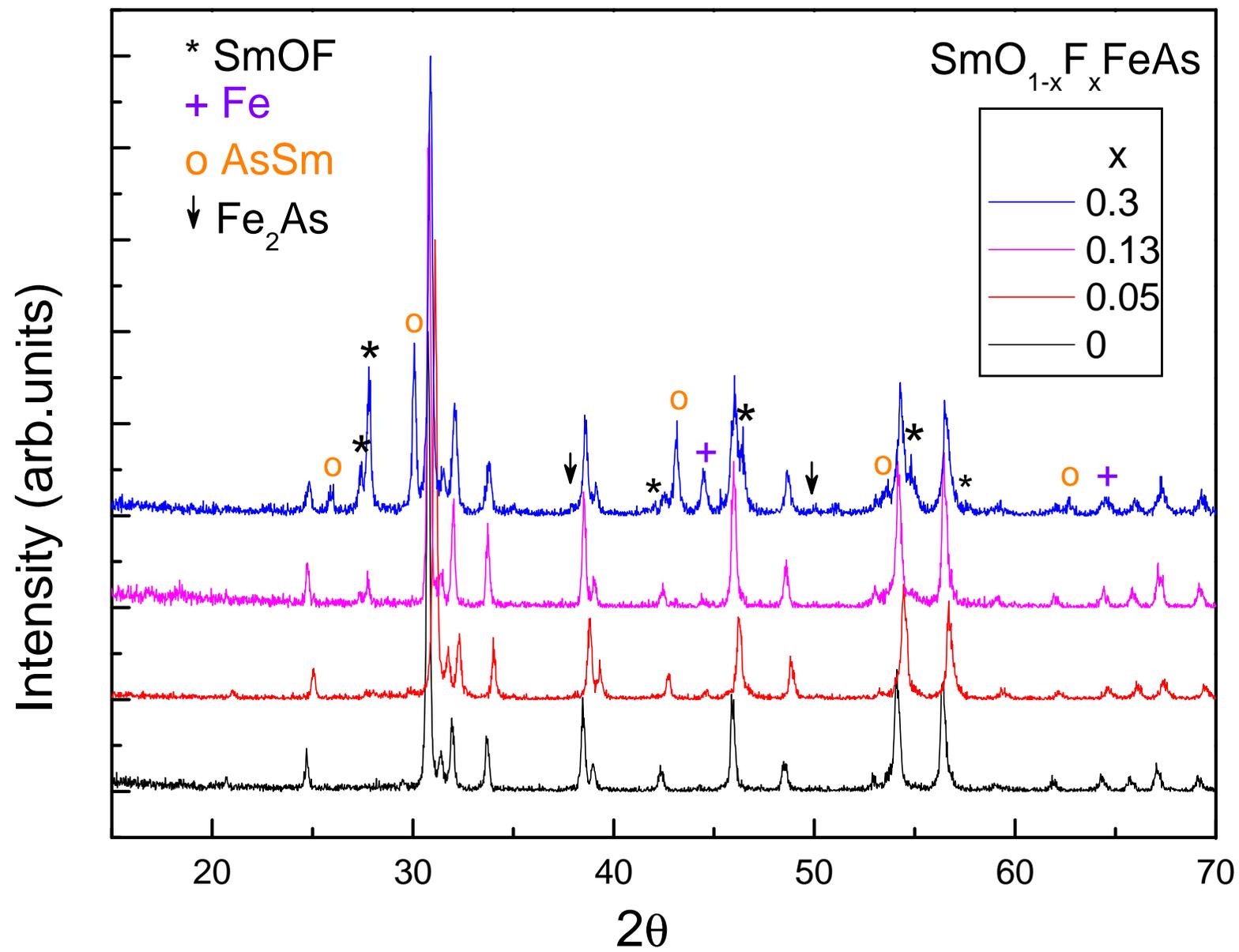
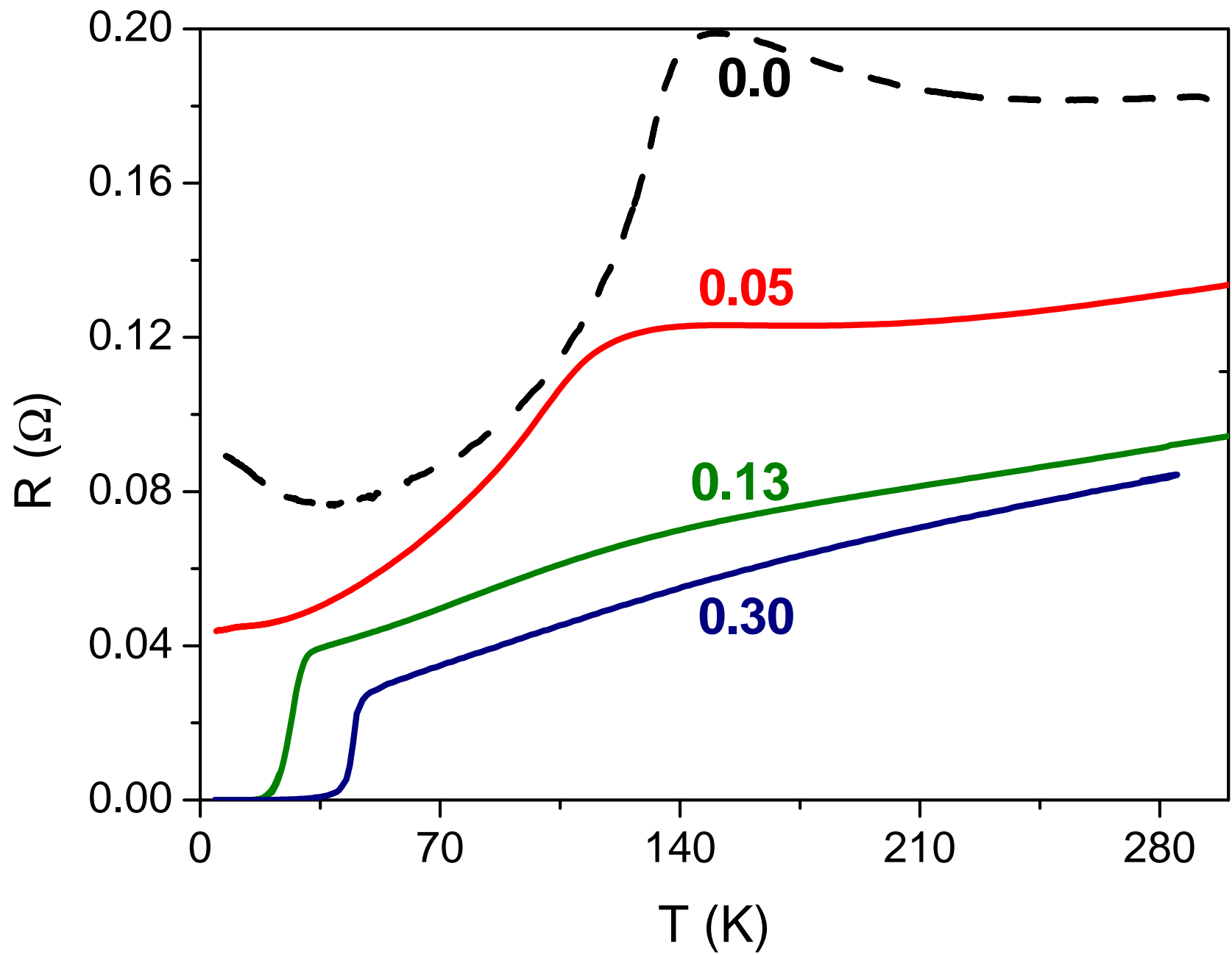
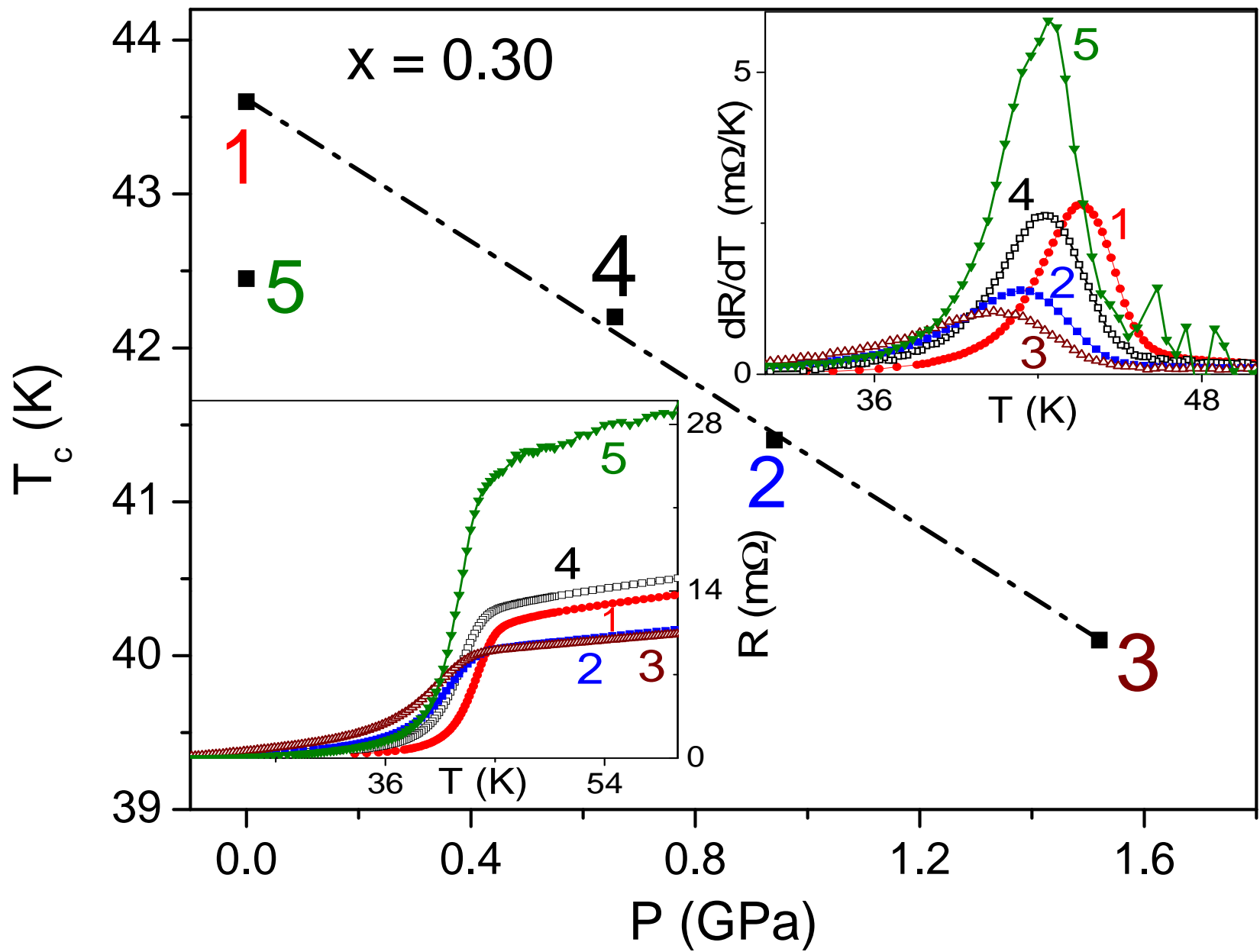
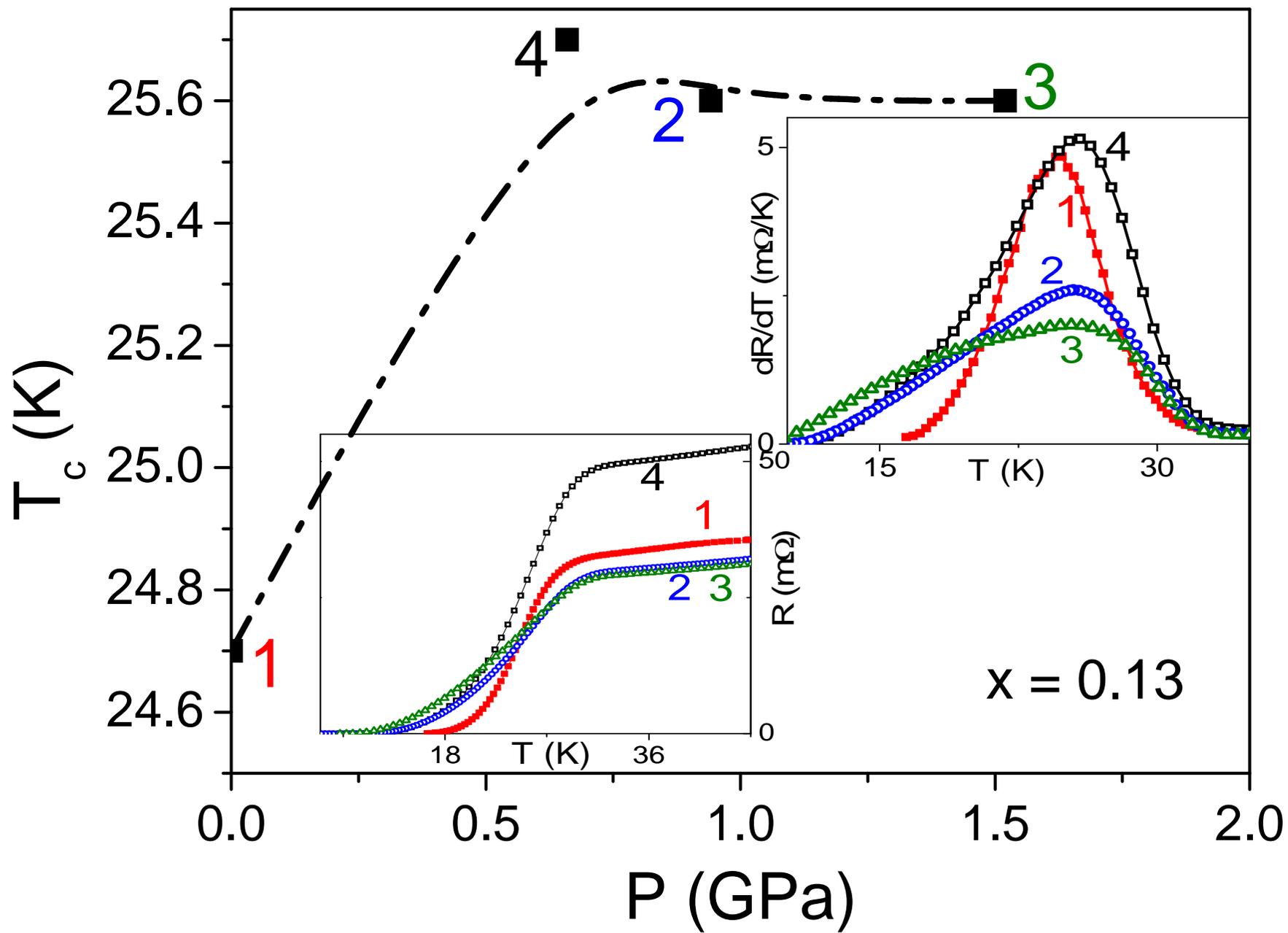
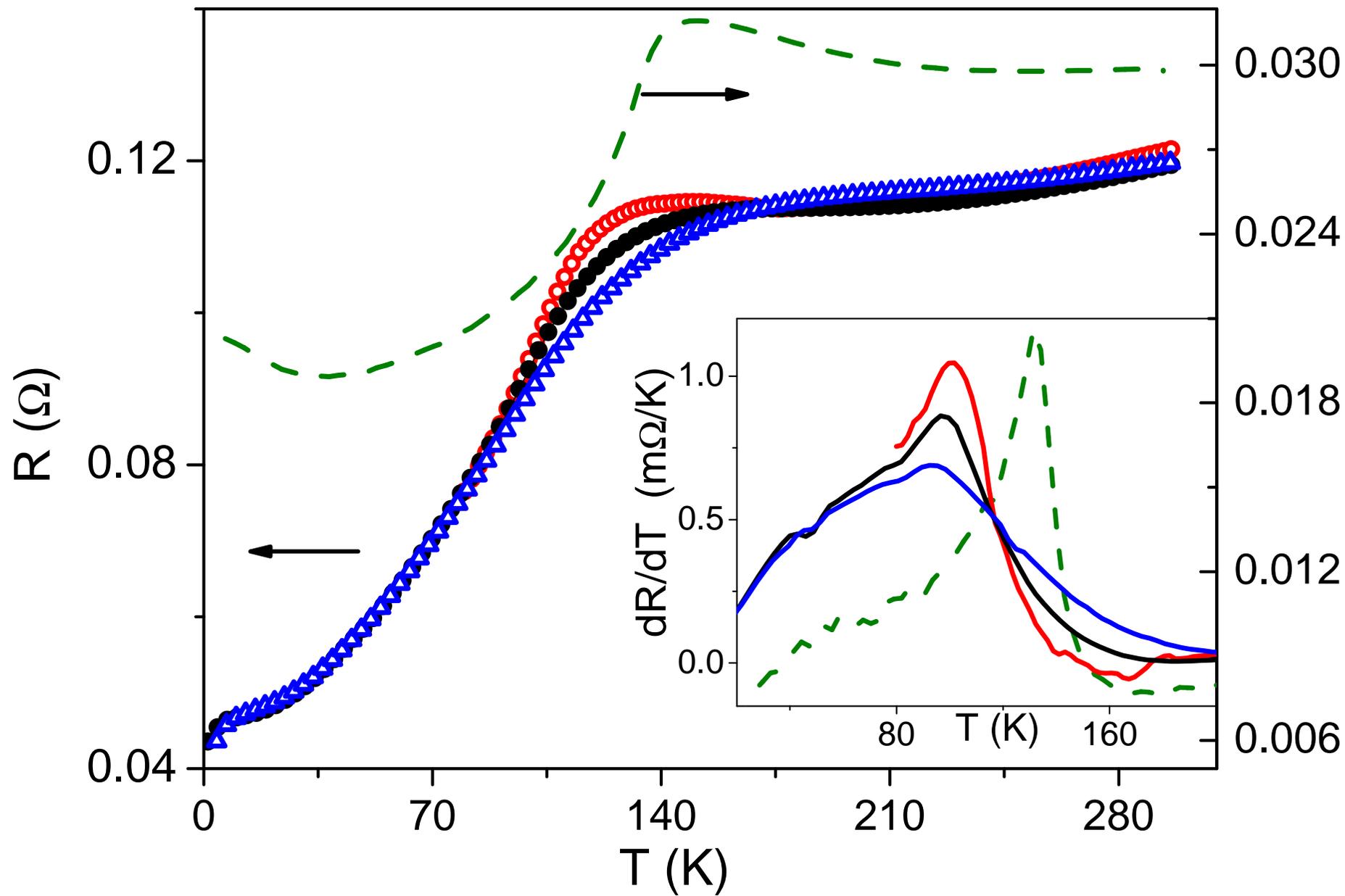
\end{document}